\documentclass[aps,prb,superscriptaddress,twocolumn,longbibliography]{revtex4-2}

\usepackage[T1]{fontenc}
\usepackage{lmodern}
\usepackage[svgnames]{xcolor}
\usepackage{lipsum}
\usepackage{graphicx}

\definecolor{Dark}{gray}{0.2}
\definecolor{MedDark}{gray}{0.4}
\definecolor{Medium}{gray}{0.6}
\definecolor{Light}{gray}{0.8}
\definecolor{darkred}{rgb}{0.55, 0.0, 0.0}
\definecolor{darkslateblue}{rgb}{0.28, 0.24, 0.55}
\definecolor{royalblue(web)}{rgb}{0.25, 0.41, 0.88}
\usepackage{graphicx,float}
\usepackage{amsmath}

\usepackage{amssymb}
\usepackage[colorlinks]{hyperref}
\hypersetup{
    citecolor=darkred,
    linkcolor=blue,   
    urlcolor=blue}

\usepackage{bbm}

\def\A{\mathrm{A}}
\def\B{\mathrm{B}}

\def\u{\uparrow}
\def\d{\downarrow}

\def\hamil{{\mathcal{H}}}
\def\lbk#1{\left(#1\right)}
\def\lmbk#1{\left[#1\right]}

\def\be{\begin{equation}}
\def\ee{\end{equation}}
\def\beq{\begin{equation}}
\def\eeq{\end{equation}}
\def\bea{\begin{eqnarray}}
\def\eea{\end{eqnarray}}
\def\nbea{\begin{eqnarray*}}
\def\neea{\nonumber\end{eqnarray*}}
\def\bmat#1{\left(\begin{array}{#1}}
\def\emat{\end{array}\right)}
\def\bcase#1{\left\{\begin{array}{#1}}
\def\ecase{\end{array}\right.}
\def\bmini#1{\begin{minipage}{#1\textwidth}}
\def\emini{\end{minipage}}

\usepackage{wrapfig}
\graphicspath{{figures/}}
\usepackage{enumerate}

\begin{document}

\title{Reviving Majorana zero modes in the spin-1/2 Kitaev ladder model}
\date{\today}
\author{Haoting Xu}
\affiliation{Department of Physics, University of Toronto, 60 St. George St., Toronto, Ontario, Canada M5S 1A7}
\author{Hae-Young Kee}
\email[]{hy.kee@utoronto.ca}
\affiliation{Department of Physics, University of Toronto, 60 St. George St., Toronto, Ontario, Canada M5S 1A7}
\affiliation{Canadian Institute for Advanced Research, CIFAR Program in Quantum Materials, Toronto, Ontario, Canada M5G 1M1 }

\begin{abstract}
The one-dimensional $p$-wave superconductor, characterized by boundary Majorana modes, has attracted significant interest owing to its potential application in topological quantum computation.  Similarly, spin-1/2 Kitaev ladder systems with bond-dependent Ising interactions, featuring Majorana fermions coupled with $Z_2$ flux, exhibit boundary Majorana modes when in a topological phase.  However, the ground state degeneracy, inherent in these systems, may result in the annihilation of Majorana modes due to the superposition of the degenerate states. To avoid this issue, here we introduce a projective measurement that selects one of the degenerate $Z_2$ sectors, enabling the revival of Majorana modes.  Once the state is selected, we show that the application of the local spin operators on a bond flips the sign of the adjacent $Z_2$ flux. Repeating such operators enables the system to reach a desired $Z_2$ flux configuration. The Majorana zero modes can be manipulated and fused by tuning the flux sectors achievable through applying local spin operators. We also discuss the engineering of the Kitaev ladder and open questions for future studies.
\end{abstract}

\maketitle

\section{Introduction\label{intro}}

\begin{figure}
    \centering
    \includegraphics{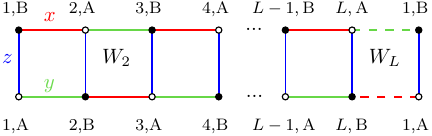}
    \caption{The Kitaev ladder with bond-dependent nearest neighbour Ising interaction. The red, green and blue bonds indicates $x$,$y$ and $z$ bond, respectively. The dashed lines denote interactions under PBC. Under OBC, the dashed lines are absent. }
    \label{fig:Kitaev_ladder}
\end{figure}

Decoherence of quantum states is recognized as one of the most serious challenges to realizing quantum computers. Topological quantum computation (TQC) provides an elegant solution to the decoherence issue by storing and manipulating quantum information non-locally in topological qubits \citep{kitaev2003fault,freedman1998p,sarma2005topologically,nayak2008non,freedman2003topological,stern2013topological}. The topological qubits are based on fractionalized excitations of many-body systems with long-range entanglement, which encode non-Abelian statistics. 

Searching for material platforms to realize fractionalized excitations has been one of the central themes in condensed matter physics. Examples of such platforms include chiral $p+ ip$ superconductors. Within the vortex core of these chiral $p+ ip$ superconductors, the emergence of Majorana zero modes, zero energy bound states, is facilitated by the additional angular momentum arising from the superconducting pairing.\cite{volovik1999fermion}
Subsequently, it was demonstrated that half-quantum vortices, constituting half of a single vortex, harboring Majorana fermions within chiral $p$-wave superconductors exhibit non-Abelian statistics.\cite{Ivanov2001}
Due to the challenges of finding the chiral $p$-wave bulk materials, proposals have been made to engineer chiral $p$-wave superconductors by leveraging the proximity effect of spin-orbit coupling in conventional superconductors.\cite{Fu2008}

An alternative approach has also been developed. 
Quantum spin liquids in frustrated magnetic systems offer fractionalized excitations. Among them, the Kitaev spin model composed of the bond-dependent Ising interaction exhibits the Majorana fermions and $Z_2$ vortices which obey the non-Abelian statistics under the magnetic field.~\citep{kitaev2006anyons}
While there is a growing list of the spin-1/2 Kitaev candidate materials, including honeycomb Na$_2$IrO$_3$, $\alpha$-RuCl$_3$, and cobaltates\cite{witczak2014correlated,Rousochatzakis_2024,rau2016spin,challenges_Winter_2016,Winter_2017,hermanns2018physics,takagi2019concept,Motome_2020,Takayama_2021,Simon2022}, the quest for a definite example of quantum spin liquids remains unresolved.

Parallel to the search for two-dimensional (2D) candidate materials,
research also focuses on one-dimensional (1D) topological systems.
In particular, the $p$-wave topological superconductor, which hosts edge Majorana zero modes, has been extensively studied.~\citep{sato2017topological,beenakker2013search,alicea2012new,leijnse2012introduction,stanescu2013majorana,elliott2015colloquium,sarma2015majorana,frolov2020topological,qi2011topological,aasen2016milestones,alicea2011non,sarma2015majorana,bai2024probing} It was demonstrated that the Majorana zero modes fuse to vacuum and fermion sector~\citep{aasen2016milestones,alicea2011non,sarma2015majorana,bai2024probing}, and the Majorana fermions encode non-Abelian statistics when forming a wire network~\citep{alicea2011non,aasen2016milestones}. 
It was shown that the braiding of Majorana fermions can be conducted with a minimum of T-junction geometry and the fusion of Majorana fermions may be detected through a parity-to-charge conversion\cite{alicea2011non,aasen2016milestones,van2012coulomb}.  

Motivated by research into 1D $p$-wave superconductors and the Kitaev honeycomb model, we investigate the 1D version of the Kitaev honeycomb model, the spin-$1/2$ Kitaev ladder, as illustrated in Fig.~\ref{fig:Kitaev_ladder}, where $x$, $y$, and $z$ refer to the spin-1/2 Ising interaction of the form $S_i^\gamma S_j^\gamma$ with $\gamma = x,y,z$.
This model is exactly solvable using the Jordan-Wigner transformation, which effectively maps the spin-1/2 system to fermionic systems through a non-local transformation.\cite{Feng2007,degottardi2011topological,chen2008exact} 
It has been demonstrated that the anisotropic Kitaev ladder phase exhibits a symmetry-protected topological (SPT) phase characterized by a string order parameter.\cite{catuneanu2019KJ_ladder,agrapidis2019KJ_ladder,sorensen2021heart_KJ_ladder,sorensen2024K_gamma_ladder,He_Ladder,Langari2015}
The Kitaev spin ladder model, featuring Majorana fermions coupled with $Z_2$ flux, exhibits edge Majorana modes when the $Z_2$ flux is fixed~\cite{degottardi2011topological,inhomo2012}. Similar to the 1D $p$-wave superconductor, previous studies propose the fusion and braiding of the Majorana zero modes using the 1D Kitaev ladder systems~\cite{inhomo2012,He2013MF,degottardi2011topological}. 
However, the $Z_2$ flux degeneracy in the ground states, inherent in the spin model, result in the annihilation of Majorana modes due to linear superposition of the ground states. 

To rescue the Majorana boundary mode, 
we introduce a projective measurement (PM) that selects one of the degenerate $Z_2$ sectors, enabling the revival of Majorana zero modes. Once the $Z_2$ sector is selected, we show the $Z_2$ sector can be manipulated by spin flips on a bond. This enables the change of topological nature by changing the flux configurations. We show that the boundary Majorana modes, together with the Majorana zero modes that reside at the interfaces of topological and non-topological (NT) phases, can be moved and fused by changing flux configurations. We also discuss the engineering of the Kitaev ladder in a material and open questions.

The paper is organized as follows. In Sec.\ref{sec:review}, we review the previous studies of the Kitaev ladder system, including the phase diagram, and the ideas of manipulating Majorana fermions. In Sec.\ref{sec:PM}, we confirm the flux sector degeneracy using numerical simulations. We show the 
absence of the boundary Majorana mode without selecting a $Z_2$ flux sector by numerical simulations. We then introduce a projective measurement to select a $Z_2$ flux sector, and show that the appearance of Majorana mode. We obtain the spin flips on bonds that effectively manipulate the $Z_2$ spin flux. In Sec.\ref{sec:manipulating}, we demonstrate that the Majorana modes can be moved and fused by changing the flux via applying the local spin operators. In Sec.\ref{sec:material}, we discuss engineering the Kitaev ladder out of the honeycomb structure by replacing $J_{\rm eff}=1/2$ ions with nonmagnetic ions except for the two zig-zag chains. 
The shape of the ladder, consisting of two coupled zig-zag chains, differs from the regular ladder due to an alternating bond length between the chains.
The phase diagram of the modified Kitaev interaction is presented. In Sec.\ref{sec:summary}, we summarize our results and discuss open questions for future studies.

\section{Brief review of Kitaev ladder and the question\label{sec:review}}
In this section we offer a brief review on the Kitaev ladder system including the phase diagram and the ideas of manipulating Majorana fermions in the Kitaev ladder~\cite{degottardi2011topological,inhomo2012,He2013MF}. However, as we will show in Sec.~\ref{sec:PM}, Majorana fermions do not exist in the Kitaev ladder systems unless a specific flux sector is selected. 

\begin{figure}
    \centering
    \includegraphics[width=0.5\textwidth]{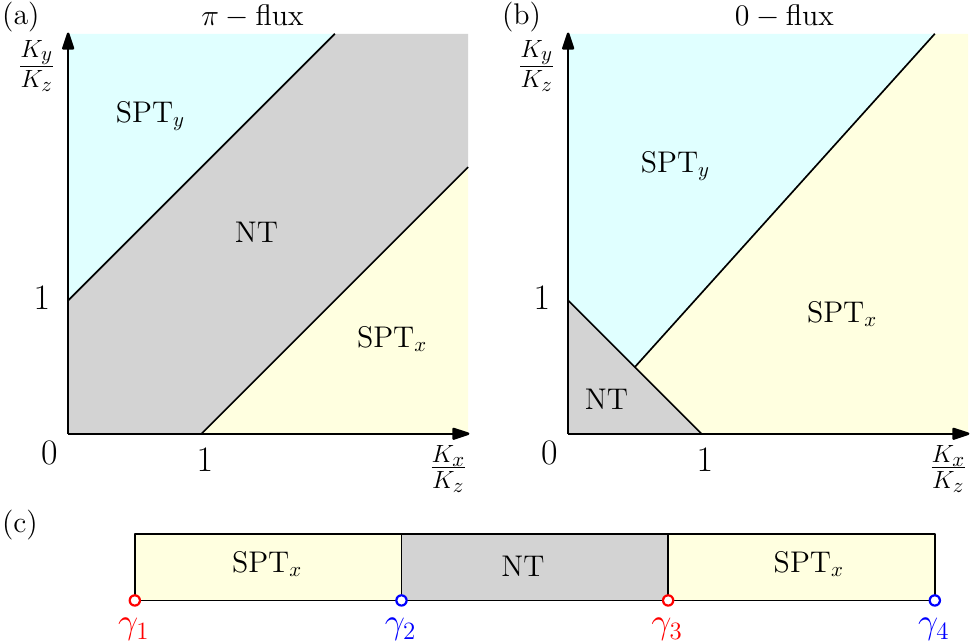}
    \caption{Brief review of previous studies on the Kitaev ladder. (a) The phase diagram of the Kitaev ladder for the $\pi$-flux (ground state) sector. (b) The phase diagram of the Kitaev ladder for the $0$-flux sector. (c) The existence of isolated Majorana mode, $\gamma$, at the intersection of the $\mathrm{SPT}_x$ phase and the NT phase after projecting the state onto a specific $\{D_j\}$ sector. The setup can be achieved through inhomogeneous ladder~\cite{inhomo2012} or tuning the flux sector (See Sec.~\ref{sec:manipulating}).
    When consider the linear combination of $\{D_j\}$ and $\{-D_j\}$ sectors, the Majorana zero modes at each end become a free complex fermion. }
    \label{fig:review}
\end{figure}

The spin-$\frac{1}{2}$ Kitaev ladder consists of bond-dependent Ising interaction with $L$ unit cells, corresponding to $N = 2L$ sites in total, as depicted in Fig. \ref{fig:Kitaev_ladder}.
The Hamiltonian of the system under open boundary condition (OBC) is given by
\begin{equation}
\label{eq:Kitaev_Hamiltonian}
    \hamil = \sum_{j=1}^{L-1} \frac{K_x}{4}\sigma_{j,\B}^x \sigma_{j+1,\A}^x + \frac{K_y}{4} \sigma_{j,\A}^y \sigma_{j+1,\B}^y
    +\sum_{j=1}^L \frac{K_z}{4}\sigma_{j,\A}^z \sigma_{j,\B}^z,
\end{equation}
where $\sigma_{j,\mu}$ are the Pauli matrices on each site and $\mu=(\A,\B)$ is the sublattice index. There exists local conserved flux defined on a square plaquette, $W_j = -\sigma_{j,\A}^x \sigma_{j,\B}^x\sigma_{j,\B}^y \sigma_{j+1,\A}^y$, as depicted in Fig.~\ref{fig:Kitaev_ladder}. The ground states of the model have $W_j = -1$ for $j$ from $1$ to $L$, which is called $\pi$-flux or vortex-full sector. This is verified by previous numerical study.~\cite{flux_gap}

The phase diagram of the model can be obtained by rewriting the Hamiltonian into a fermion model, via Jordan-Wigner transformation (JWT) or the four Majorana representation~\cite{kitaev2006anyons}. In this paper we use JWT, where a spin-1/2 on each site, $\sigma_{j,\mu}$, is represented by two Majorana fermions, $\gamma_{j,\mu}$ and $\Tilde{\gamma}_{j,\mu}$. Hence, JWT preserves the size of the Hilbert space. 
In the four Majorana representation, it is not obvious to see the flux degeneracy of the spin model as described below. 
The detailed definition of JWT is reviewed in Appendix~\ref{append:JW}.
In the Majorana fermion representation, the ladder Hamiltonian is represented by
\begin{equation}
\label{eq:K_ladder_fermion}
\begin{aligned}
    \hamil =& \sum_{j=1}^{L-1} \left( \frac{K_x}{4} i\gamma_{j,\B}\gamma_{j+1,\A}- \frac{K_y}{4} i\gamma_{j,\A}\gamma_{j+1,\B}\right) \\
    &+ \sum_{j=1}^L\frac{K_z}{4} (i\gamma_{j,\A}\gamma_{j,\B})(i\Tilde{\gamma}_{j,\A}\Tilde{\gamma}_{j,\B}).
\end{aligned}
\end{equation} 
We define a $Z_2$ flux on the vertical bond, $D_j$, as $D_j = i\Tilde{\gamma}_{j,\A}\Tilde{\gamma}_{j,\B}$, which commutes with the Hamiltonian and each other.  Note that there is no Hilbert space enlargement in this scheme as discussed above.  This should not be confused with the $D_j$ in the four-Majorana representation~\cite{kitaev2006anyons,inhomo2012}, where $D_j$ is employed to select states from an enlarged Hilbert space into the physical space. In contrast, the $D_j$ defined here is an operator within the physical Hilbert space itself.
The $Z_2$ flux, $D_j$, can take values of either $+1$ or $-1$.
Hence the Hamiltonian, $\hamil$, can be diagonalized in different $\{D_j\}$ configurations.  The $\pi$-flux sector corresponds to $W_j \equiv D_j D_{j+1} = -1$. This means that the ground state has the $\{D_j\}$ configuration of either $\{D_j\}=\{+1,-1,\cdots\}$ or $\{D_j\} = \{-1,+1,\cdots\}$. 
The sector $\{D_j\}$ and the sector $\{-D_j\}$ have the same energy spectrum for Majorana fermions. This is because the transformation $D_j \to -D_j$ for all $j$ is equivalent to $\gamma_{j,\mu} \to -\gamma_{j,\mu}$ for $(j,\mu) \in \mathcal{L}_B$, where $\mu=A, B$ and
$\mathcal{L}_B$ and $\mathcal{L}_T$ denote the bottom and the top chain of the ladder, respectively. Such a local phase factor added to the definition of Majorana fermions does not change the energy spectrum. 

The phase diagram, edge Majorana modes and methods for manipulating the Majorana modes are studied, assuming a quadratic Majorana fermion Hamiltonian, i.e., fixed $\{D_j\}$ sectors.~\cite{Feng2007,inhomo2012,degottardi2011topological,He2013MF,math_paper_2013,math_paper_2015} 
To obtain the phase diagram of the Kitaev spin ladder for a given flux sector, we exactly solve the quadratic Hamiltonian for the Majorana fermions for given $\{D_j\}$.
In a given $\{D_j\}$ sector, there are four bands in the dispersion, originating from the four sublattices. The ground state of the model corresponds to filling all the bands below zero energy. 
The energy dispersion of the four bands takes the form of $(\epsilon_{k,+}^{\pi}, \epsilon_{k,-}^{\pi},-\epsilon_{k,+}^{\pi}, -\epsilon_{k,-}^{\pi})$, where
\begin{equation}
    \epsilon_{k,\pm}^{\pi} =  \frac{1}{2} \sqrt{(K_z\pm (K_x - K_y)\sin \frac{k}{2})^2 + (K_x+K_y)^2 \cos^2 \frac{k}{2}}.
\end{equation}
The ground state energy is the sum over the two occupied bands, 
\begin{equation}
    E_{gs} = -\sum_{k=-\pi}^\pi (\epsilon_{k,+}^{\pi}+\epsilon_{k,-}^{\pi}). 
\end{equation}
The energy gap closes at $\vert K_x - K_y \vert = K_z$, indicating a topological phase transition at these lines. For $0$-flux sector, i.e., $\{D_j\} = \{+1,+1,\cdots\}$ or $\{D_j\} = \{-1,-1,\cdots\}$, the fermion bands are $(\epsilon_{k,+}^{0}, \epsilon_{k,-}^{0},-\epsilon_{k,+}^{0}, -\epsilon_{k,-}^{0})$, where
\begin{equation}
    \epsilon_{k,\pm}^{0} = \frac{1}{2} \sqrt{(K_z\pm (K_x + K_y)\cos\frac{k}{2})^2+(K_x - K_y)^2 \sin^2 \frac{k}{2}}.
\end{equation}
Here, the transition lines are $\vert K_x + K_y \vert = K_z$ and $K_x = K_y$ with $(\vert K_x + K_y\vert>K_z)$. The phase diagram of the $\pi$-flux sector and the $0$-flux sector is depicted in Fig.~\ref{fig:review}(a) and (b). The phases are identified as SPT$_x$, SPT$_y$, and non-topological (NT) phase, respectively. By assuming a fixed flux configuration, the Majorana edge modes are identified, in analogous to the Kitaev chain.~\cite{degottardi2011topological} 
The fusion and braiding of the Majorana fermions at the intersection of the NT and SPT phases in the ladder systems as shown in Fig. ~\ref{fig:review}(c) have also been proposed using an inhomogeneous ladder system~\cite{inhomo2012} and quench dynamics~\cite{He2013MF}.

\section{Projective measurement\label{sec:PM}}
\begin{figure}
    \centering
    \includegraphics[width = 0.5\textwidth]{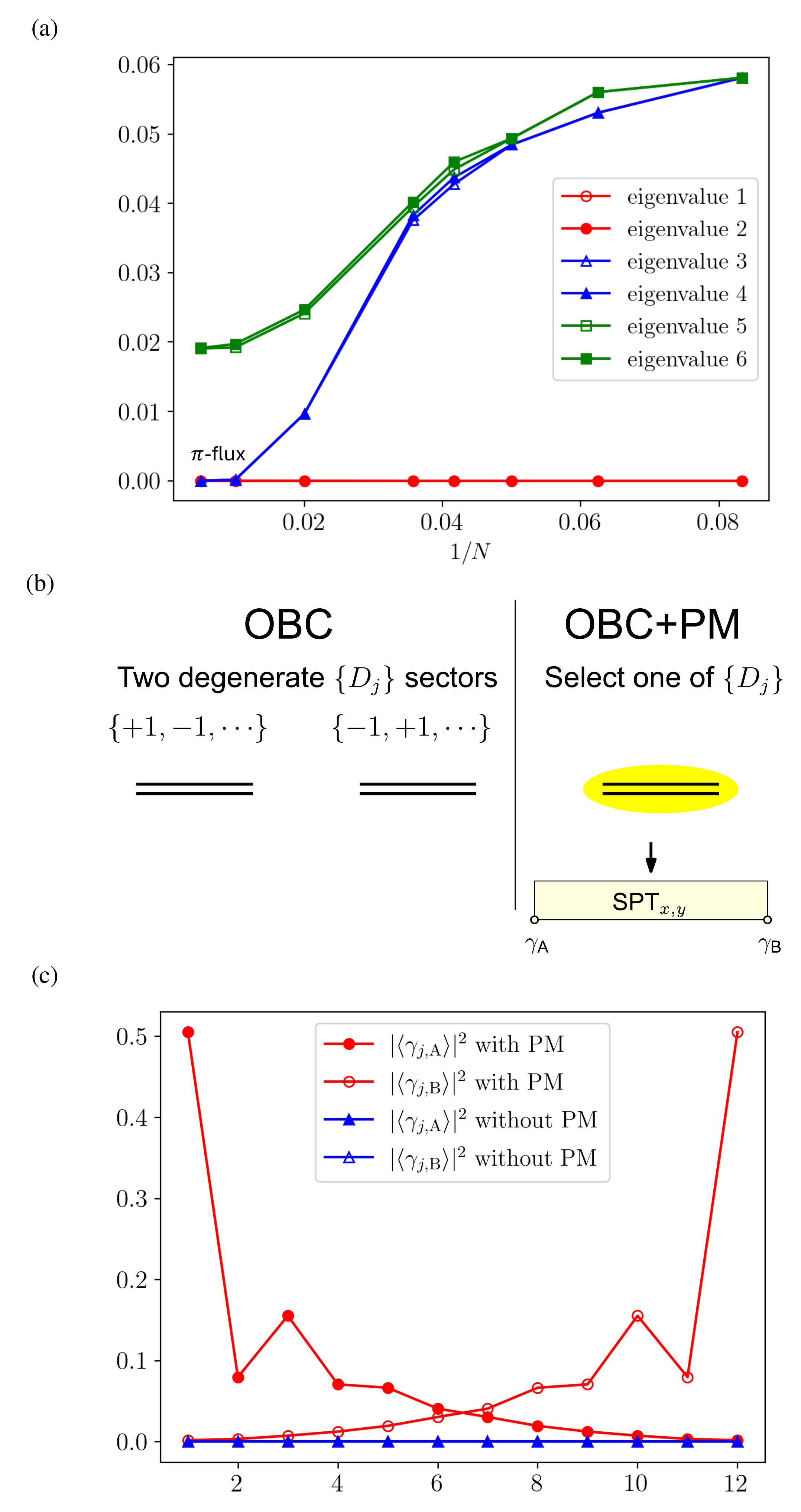}
    \caption{(a)The energy difference with the ground state of the six lowest eigenstates, obtained from ED (for $N=12,16,20,24,28$) and DMRG (for $N=50,100,200$). As the system size increase, the 4-fold ground state degeneracy is clear. It is also confirmed numerically that the ground state has $\pi$-flux.(b) 
    The 4-fold degenerate ground states in the SPT$_{x/y}$ phase for OBC. Two-fold degeneracy are from different $\{D_j\}$ sectors. Within each $\{D_j\}$ sector, there exists Majorana edge modes at the ends of the ladder, resulting in another 2-fold degeneracy. (c) The probability of localized Majorana fermions for ground states of the Kitaev ladder with the system size $N=24$. The blue lines represent the ground states without PM, and the red lines represent the ground states after PM indicating the absence of the boundary modes without PM.} 
    \label{fig:degeneracy}
\end{figure}

In this section we first verify the 4-fold degeneracy in the SPT$_x$ phase using numerical simulations. The 4-fold degenerate ground states can be understood as two different $\{D_j\}$ sectors, with each of the sector containing two boundary Majorana modes. We propose the PM to obtain the desired ground states with edge Majorana modes. We demonstrate that without PM, the Majorana edge modes disappear. 

In the SPT$_x$ phase, when the $\{D_j\}$ configuration is fixed, two Majorana modes are located at the ends of the ladder, which are characterized by the Majorana fermion operators, $\gamma_\A$ and $\gamma_\B$. For a fixed $\{D_j\}$ configuration, the two degenerate ground states can be characterized by eigenstates of either $\gamma_\A$ or $\gamma_\B$. See Appendix~\ref{appendix:boundary_mode} for details.
However, the Majorana representation does not reflect the physics of the spins properly. 
In fact, for the isotropic Kitaev ladder $(K_x=K_y=K_z)$ in the NT phase, the ground state has $4$-fold degeneracy when considering a different OBC.~\cite{sorensen2024K_gamma_ladder} In the SPT$_{x/y}$ phase, the 4-fold ground state degeneracy is confirmed recently by identifying these phases as $Z_2\times Z_2$ SPT phase.~\cite{He_Ladder} The low energy effective Hamiltonian in the SPT$_{x/y}$ phase has the form of the ZXZ cluster model, which is connected to the Haldane phase under a local unitary transformation~\cite{1DSPTs}. Equivalently, the spin system contains two degenerate $\{D_j\}$ sectors, where the states from these two sectors are orthogonal and degenerate. The flux sector degeneracy also exists in the NT phase, which is presented in Appendix~\ref{sec:perturbation_theory}.

To confirm the ground state degeneracy, 
we perform exact diagonalization (ED) and density matrix renormalization group (DMRG) of the Kitaev ladder in the SPT$_x$ phase with $K_x = 2.5, K_y=K_z=1.0$.
The 6 lowest eigenvalues are found and depicted in Fig.~\ref{fig:degeneracy}(a). As the system size increases, the blue and red lines join together, indicating a 4-fold ground state degeneracy. In Fig.~\ref{fig:degeneracy}(b), we illustrate the reason for the 4-fold degeneracy in OBC. There are two degenerate $\{D_j\}$ sectors, and  there are two-fold degeneracy due to the edge Majorana modes in each sector. If one does not select $\{D_j\}$, i.e., taking an arbitrary linear combination of the four ground states, one cannot obtain the edge Majorana modes.

\begin{figure}
    \centering
    \includegraphics[width = 0.45\textwidth]{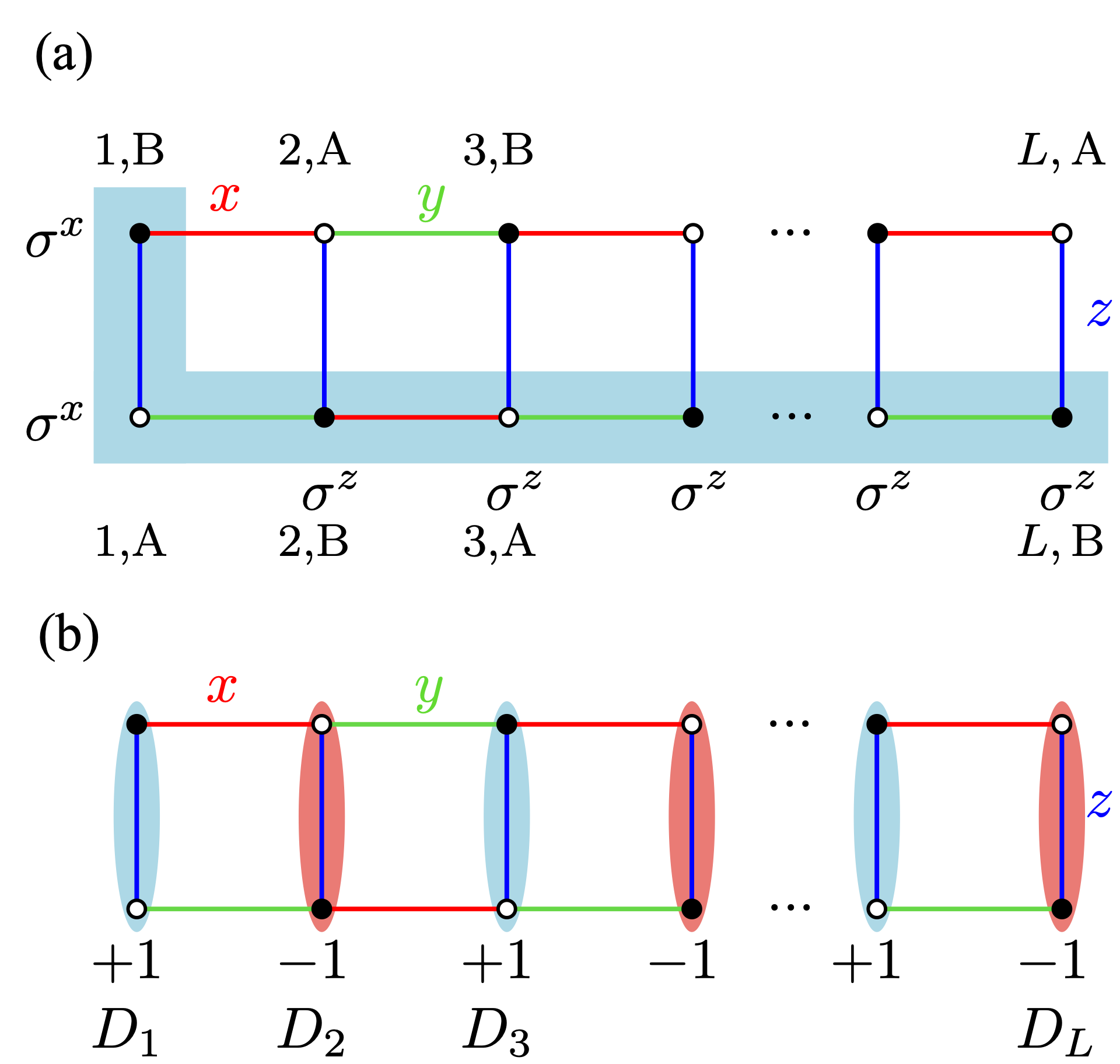}
    \caption{(a) Illustration of the $D_1$ operator given in the main text which involves a series of spin operators on the sites along blue shaded L shape. 
    (b) Flux pattern of one of the ground state selected by PM with $\{D_j\}=\{+1,-1,\cdots\}$.
    The blue shaded vertical bonds represent $D_j=+1$, and the red shaded vertical bonds represent $D_j = -1$. 
    }
    \label{fig:Dj}
\end{figure}

In order to obtain the edge Majorana modes in the topological phase, SPT$_x$ or SPT$_y$, the $\{D_j\}$ degeneracy must be broken by choosing a specific flux sector. 
We propose that this can be done by applying the following projective measurement of $D_j = i\Tilde{\gamma}_{j,\A}\Tilde{\gamma}_{j,\B}$, which corresponds to non-local spin operations in spin space.
We first write $D_j$ operators in the spin representation, 
\begin{equation}
    D_j = \left \{
    \begin{aligned}
        &\sigma_{j,\A}^x \sigma_{j,\B}^x\prod_{\substack{k>j\\(k,\mu)\in \mathcal{L}_B}} \sigma_{k,\mu}^z \prod_{\substack{l<j\\ (l,\nu)\in \mathcal{L}_T}} \sigma_{l,\nu}^z, \;\; j \text{ is odd,}\\
        &-\sigma_{j,\A}^y \sigma_{j,\B}^y \prod_{\substack{k>j\\ (k,\mu)\in \mathcal{L}_B}} \sigma_{k,\mu}^z \prod_{\substack{l<j\\ (l,\nu)\in \mathcal{L}_T}} \sigma_{l,\nu}^z, \;\; j\text{ is even.}
    \end{aligned}
    \right.
    \label{eq:Dj_spin}
\end{equation}
Since the ground states always satisfy $D_jD_{j+1}=-1$, the PM can be achieved by only applying 
$\Pi_j = (1+(-1)^{j+1}D_j)/2$ for any specific $j$.
Here we demonstrate applying $\Pi_1 = (1+D_1)/2$ as an example.  In terms of the spin operators, $D_1$ is given by 
\begin{equation}
    D_1 = \sigma_{1,\A}^x \sigma_{1,\B}^x \prod_{(j,\mu)\in \mathcal{L}_B}\sigma^z_{j,\mu},
\end{equation}
which is a series of spin operators acting on sites along an ``L'' shape, as depicted in Fig.~\ref{fig:Dj}.
Considering any linear combination of the ground states, which is written as $\vert \psi\rangle = \psi_1 \vert \{+1,-1,\cdots\}\rangle  + \psi_2 \vert \{-1,+1,\cdots\}\rangle $, the $\Pi_1$ operator selects the desired $\{D_j\}=\{+1,-1,\cdots\}$ sector, and sends the other term to zero. 
One can choose $D_j$ for any other $j$-bond
instead of $D_1$.

Once the state is chosen by applying PM, $\Pi_1$ operator, the $Z_2$ flux can be manipulated by local spin operators.  
The local spin operators that flip the sign of $D_j$ and $D_{j+1}$ are
\begin{equation}
U_{(j,j+1)} = \sigma_{j,\A}^x \sigma_{j+1,\B}^x = i \Tilde{\gamma}_{j,\A} \Tilde{\gamma}_{j+1,\B},
\end{equation}
which are nearest-neighbor spin operators. It is straight forward to check that 
\begin{equation}
    U_{(j,j+1)} D_l U^\dagger_{(j,j+1)} = -D_l (\delta_{l,j} + \delta_{l,j+1}),
\end{equation} 
which means that the operation $U_{j,j+1}$ flips the sign of $D_j$ and $D_{j+1}$.
The unitary operator can also be $\sigma_{j,\B}^y \sigma_{j+1,\A}^y = -i \Tilde{\gamma}_{j,\B} \Tilde{\gamma}_{j+1,\A}$, which flips the sign of $D_j$ and $D_{j+1}$ as well. Hence, any configuration of $\{D_j\}$ can be reached by applying projective quantum measurement and local spin operators. A potential realization of applying local spin operators on the spin systems is discussed in Appendix \ref{sec:spin_operator}. 

To verify the effects of the PM,  
we take the first two ground states (denoted by red lines) from our ED simulation and compute the probabilities of localized Majorana fermion,
 $\vert \langle \gamma_{j,\A}\rangle\vert^2$ and $\vert \langle \gamma_{j,\B}\rangle\vert^2$, for the chosen state. 
We found that 
without the PM, the Majorana modes disappear as shown in the blue lines in Fig.~\ref{fig:degeneracy}(c). 
On the other hand, there are two Majorana zero modes after PM, one is localized near the left edge and the other one is located near the right edge, as shown in the red lines of Fig.~\ref{fig:degeneracy}(c). 
One can see the exponential decay of the probability into the bulk of the system, even for a small system size, i.e., $L=12$. 
These results confirm that there exists Majorana edge modes only after performing the PM, which selects a specific $\{D_j\}$ configuration.

\section{Manipulating Majorana fermions\label{sec:manipulating}}

\begin{figure}
  \centering
    \includegraphics[width = 1.0\linewidth]{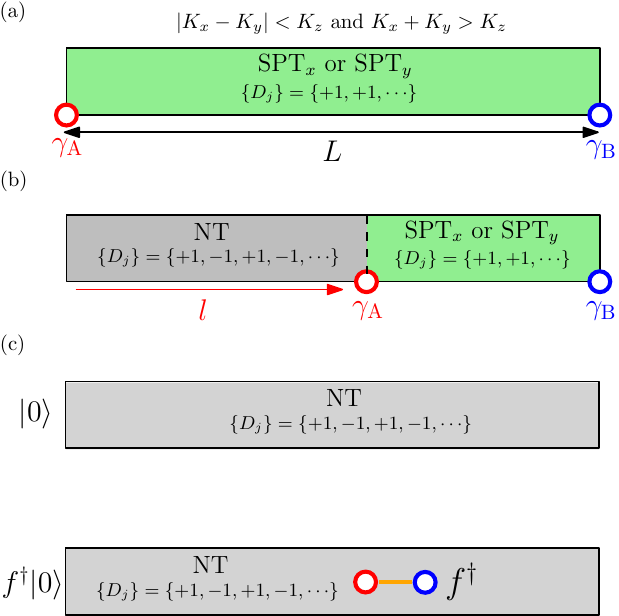}
  \caption{The illustration of moving Majorana fermion. The interaction strength is set to satisfy $\vert K_x - K_y\vert < K_z$ and $K_x+K_y >K_z$. (a)The system is initialized  with $\{D_j\} = \{+1,+1,\cdots\}$, which gives the topological phase, $\mathrm{SPT}_x$ or $\mathrm{SPT}_y$, with edge Majorana fermion $\gamma_\A$ and $\gamma_\B$.
    (b)The Majorana fermion, $\gamma_\A$, is moved by spin operations that change the $Z_2$ flux, $D_j$. The $Z_2$ flux configuration is changed to $\{+1,-1,+1,-1,\cdots\}$, which turns the gray region from topological phase to the NT phase. The Majorana fermion $\gamma_\A$ is adiabatically moved from the left boundary of the material to the interface of the NT and topological phase. (c) The two resulting state of Majorana fusion. Depending on the initial state in (a), the state after fusion is either vacuum, $\vert 0\rangle$ or a fermion excitation, $f^\dagger \vert 0 \rangle$.}
  \label{fig:moving_fusion}
\end{figure}

In this section we demonstrate how to move and fuse the Majorana fermions by changing the $\{D_j\}$ sector. From the phase diagram, Fig.~\ref{fig:review}, the SPT phases and NT phases have different region for different flux sectors. Hence, it is possible to manipulate the Majorana fermions without changing the coupling constant of the system. 

Here we give an example using the $\pi$-flux and $0$-flux phase diagram. 
In order to change the topological nature of the system through changing $\{D_j\}$ configurations, 
the parameters are chosen to satisfy $\vert K_x - K_y \vert < K_z$ and $K_x+K_y>K_z$. 
The system is initially prepared to be in SPT phase, i.e., either $\mathrm{SPT}_x$ or $\mathrm{SPT}_y$ phase, 
with $\{D_j\} = \{+1,+1,\cdots\}$, which is depicted in Fig.~\ref{fig:moving_fusion}(a). 
In the initial setup, two Majorana fermions, $\gamma_\A$ and $\gamma_\B$, reside on the left and right edge of the ladder. Here we demonstrate how to move the left Majorana fermion, $\gamma_\A$, through the change of the flux. This can be done by considering the ladder system with different parts. As shown in Appendix~\ref{appendix:boundary_mode}, an isolated Majorana mode is located at the intersection between the SPT and non-topological phases. According to Sec.~\ref{sec:PM}, one can apply $\sigma_{2,\A}^x\sigma_{3,\B}^x$ to change the flux configuration to $\{D_j\}=\{+1,-1,-1,+1,+1,+1,\cdots\}$. One then apply $\sigma_{3,\A}^x\sigma_{4,\B}^x$, the flux configuration becomes $\{D_j\} = \{+1,-1,+1,-1,+1,+1,+1,\cdots\}$. One then repeat the operations of $\sigma_{j,\A}^x \sigma_{j+1,\B}^x$ from $j=4$ to $j= l$ in sequence. Here we assume $l$ is an even number for convenience. These processes drive the system to a flux configuration consist of $\pi$-flux pattern with length $l$ and $0$-flux pattern with length $L-l$, as shown in Fig.\ref{fig:moving_fusion}(b). In this flux configuration, the system consists of two Majorana zero modes, one is located in the intersection between the NT and SPT phases, another one is the $\gamma_\B$ on the right end. Hence, one effectively move the left-end Majorana fermion, $\gamma_\A$, to the middle of the ladder. One can further move $\gamma_\A$ to the right end of the system, and this result in the fusion of two Majorana fermions, $\gamma_\A$ and $\gamma_\B$. When their distance is shorter, they interact with each other and are no longer zero modes. 
The two consequences of fusion of the two Majorana fermions are illustrated in Fig.\ref{fig:moving_fusion}(c), one is the vacuum state, and the other is a fermion excitation. 

\begin{figure}
    \centering
    \includegraphics[width = 0.9\linewidth]{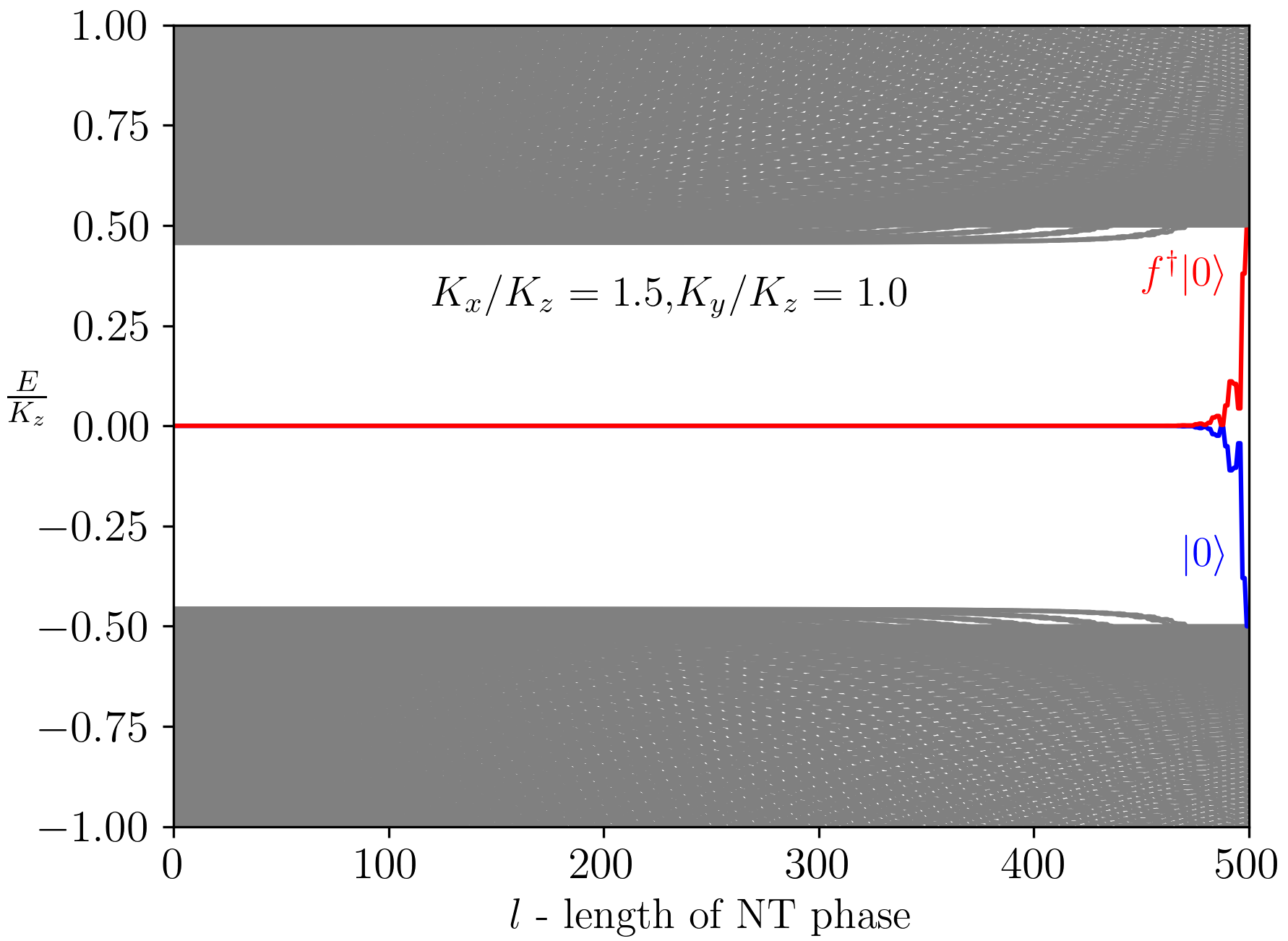}
    \caption{The numerical simulation of the moving and fusion process. The energy spectrum for $E/K_z$ from $-1$ to $1$ is plotted. After fusion the state becomes either the ground state or with a fermion excitation, indicated by the blue and red lines.}
    \label{fig:fusion_data}
\end{figure}

Numerical simulation is conducted in order to confirm the moving and fusion of Majorana fermions by changing $\{D_j\}$. Note that all the degeneracy described below is based on the system after PM. The Majorana fermion Hamiltonian with OBC is diagonalized with $N=1000$ sites with fixed parameter choice $(K_x/K_z,K_y/K_z) = (1.5,1)$. The process depicted in Fig.~\ref{fig:moving_fusion} is simulated numerically. The energy spectrum from $-K_z$ to $K_z$ is plotted as a function of $l$ in Fig.~\ref{fig:fusion_data}, where $l$ denotes the length of the NT phase. We start with $l=0$, which corresponds to Fig.~\ref{fig:moving_fusion}(a), i.e., all the material is within the SPT. For a finite $l$, we confirm the existence of the double degeneracy, which is attributed to two isolated Majorana fermions. One of the Majorana fermions resides at the interface of the NT and SPT phase, while the other Majorana fermion resides on the right edge of the system, as depicted in Fig.~\ref{fig:moving_fusion}. When $l$ approaches $L=500$, the rapid increasing of the energy of the complex fermion is observed. One of the state becomes the ground state, which is represented by the blue line. The other state becomes a complex fermion excitation, $f=\frac{1}{2}(\gamma_\A+i \gamma_\B)$, which has finite energy in the Hamiltonian.
 
\section{Engineering a Kitaev spin ladder and challenges\label{sec:material}}
\begin{figure}
    \centering
    \includegraphics[width = 1.0\linewidth]{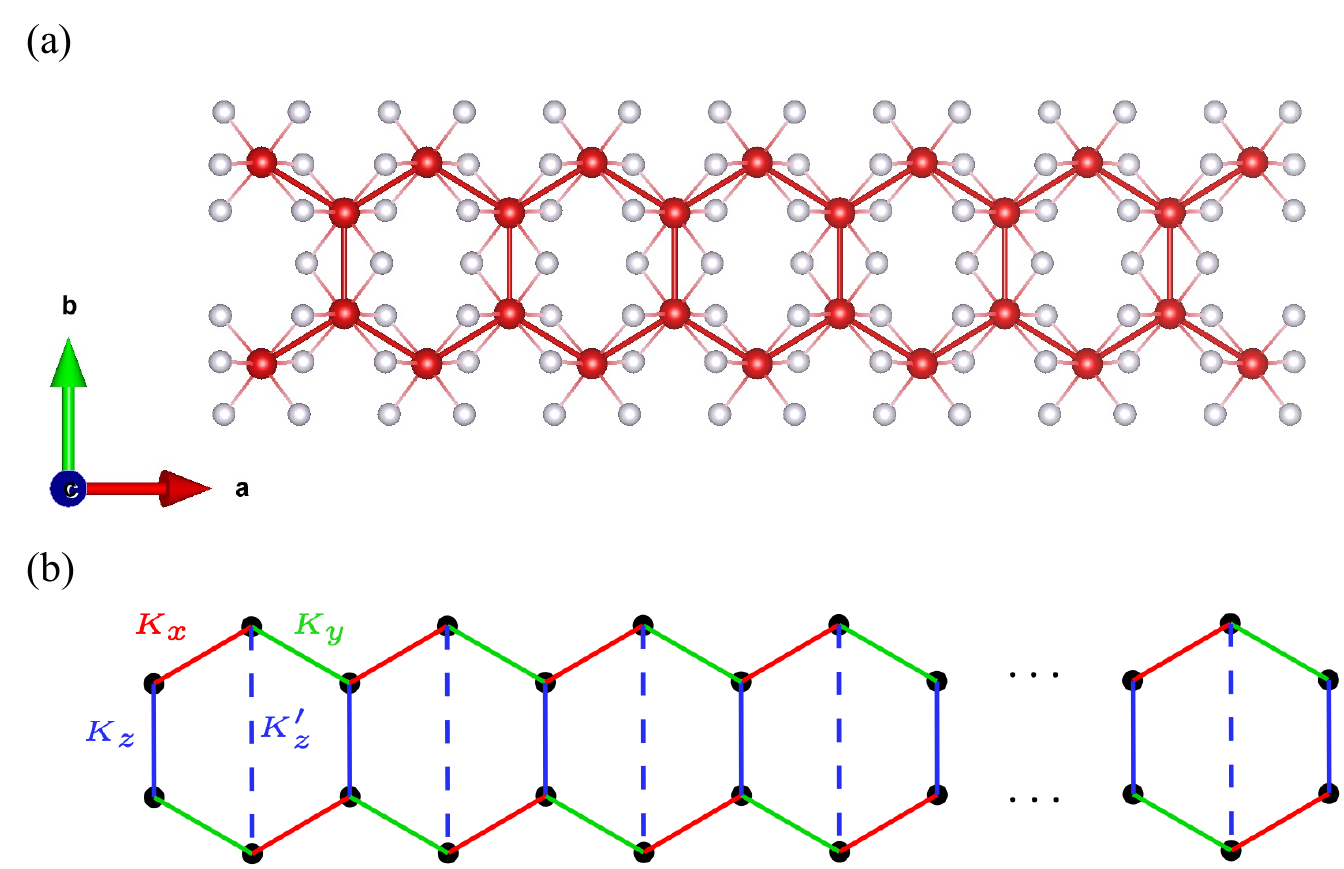}
    \caption{(a) An illustration of the engineered Kitaev ladder. The red dots represent atoms with $J_{\rm eff}$=1/2 wavefunction surrounded by $p$-orbitals of ligand atoms depicted by white dots. The ladder is confiend by the insulating barrier. See the main text for details. (b) The trimmed ladder geometry with the alternating $z$-bond interaction. $K^\prime_z$ represents the $z$-bond with a longer distance. } 
    \label{fig:material}
\end{figure}

In this section we propose how to engineer a Kitaev spin ladder. The Kitaev interaction is originated from the indirect exchange process between $J_{\rm eff}=1/2$ wavefunction at the transition metal site and $p$-orbital at anion sites when they make $90^\circ$ bond angle.\cite{Jackeli2009}  
For example, for Ru$^{3+}$ ions with 4d$^5$ electron configuration surrounded by Cl$^-$ octahedron, the low energy physics is described by $J_{\mathrm{eff}}=1/2$, due to a combination of electron correlation and spin-orbital coupling. The bond-dependent Kitaev term is emerged from the interactions between these $J_{\mathrm{eff}}=1/2$ pseudo-spins via indirect exchange path with $p$-orbital. In order to get the Kitaev ladder geometry as illustrated in Fig. \ref{fig:material}(a), the exchange path towards the adjacent ladders must be frozen. To achieve this, we replace the magnetic $J_{\rm eff}=1/2$ ions, for instance RuCl$_3$, with non-magnetic ions such as IrCl$_3$ with filled $d^6$ ions, except for the two coupled zig-zag chains as shown in Fig. ~\ref{fig:material}(a).
To block a possible charge transfer between the magnetic and nonmagnetic ions, a thin insulating barrier may be needed.

Note that Fig.~\ref{fig:material}(a) is merely an illustration rather than a real crystal structure. The exchange interaction between sites would give us the bond-dependent Kitaev interaction, as well as other interactions. 
The 1D Kitaev-like term can be realized in this trimmed honeycomb ladder through the nearest-neighbor interaction, $K = (K_x,K_y,K_z)$ as discussed above, but one of the $z$-bond interactions with a longer distance is replaced by $K^\prime_z$, as depicted in Fig. ~\ref{fig:material}(b).  

\begin{figure}[t]
\centering
    \includegraphics[width = 1.0\linewidth]{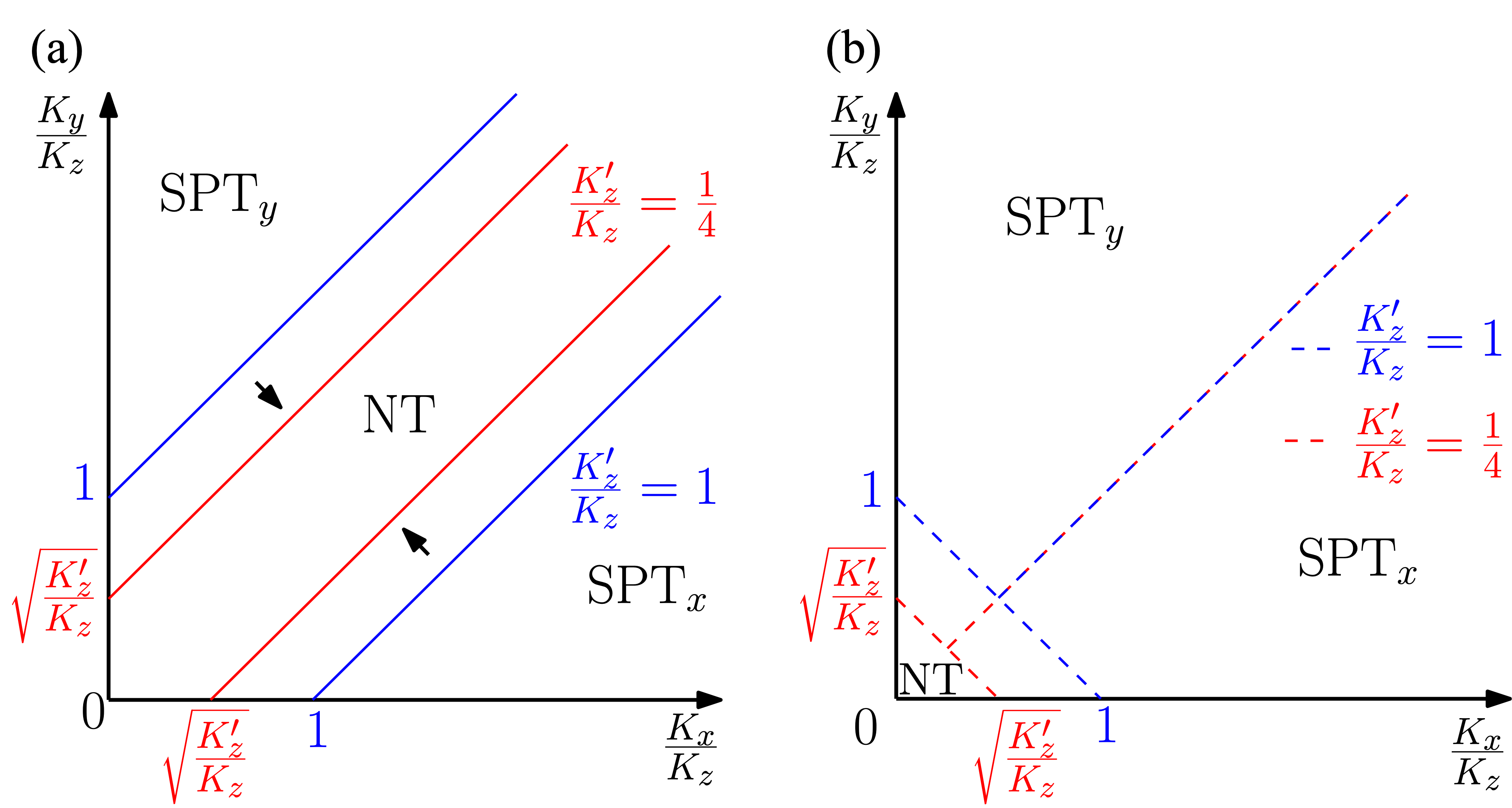}
    \caption{Phase diagram of the trimmed honeycomb ladder. Blue and red color represent $K^\prime_{z}=K_{z}$, the original model studied in previous section, and $K^\prime_{z} =\frac{1}{4} K_{z}$, respectively. The black arrows show the change of the transition lines when $K^\prime_{z}/K_{z}$ is moving from $1$ to a smaller value. (a) Solid lines represent transition line in the ground state sector, $\{D_j\} = \{+1,+1,\cdots\}$, i.e., $\pi$-flux. (b) Dashed lines represent transition lines in the $0$-flux sector, where $\{D_j\}=\{+1,-1,+1,-1,\cdots\}$.  } 
    \label{fig:pd_realistic}
\end{figure}

Taking into account the alternating $z$-bond interaction for the trimmed honeycomb ladder, and 
writing the spin-spin interaction in terms of Majorana fermions, which have the same definition in Sec.~\ref{sec:review}, the Hamiltonian reads:
\begin{equation}
\begin{aligned}
    \hamil =& \frac{1}{4} \sum_{j=1}^{L} [i K_{x} \gamma_{2j-1,\B}\gamma_{2j,\A} - i K_{y} \gamma_{2j-1,\A}\gamma_{2j,\B} \\
    &+K_{z} (i\gamma_{2j-1,\A}\gamma_{2j-1,\B})D_{2j-1} \\
    &+ i K_{x} \gamma_{2j,\B}\gamma_{2j+1,\A} - i K_{y}\gamma_{2j,\A}\gamma_{2j+1,\B}\\
    &+ K^\prime_{z} (i\gamma_{2j,\A}\gamma_{2j,\B})D_{2j}].\\
\end{aligned}
\end{equation}
We numerically check that the ground state resides in the $\pi$-flux sector. In the $\pi$-flux sector, the energy dispersion has four bands, $(\epsilon_{k,+},\epsilon_{k,-},-\epsilon_{k,+},-\epsilon_{k,-})$, where
\begin{equation}
\begin{aligned}
    &\epsilon_{\pm, k}^2 = (K_{x} + K_{y})^2 \cos^2 \frac{k}{2} + (K_{x} - K_{y})^2 \sin^2 \frac{k}{2} \\
    &+ \frac{1}{2}(K_{z}^2 + K^{\prime 2}_{z}) \\ 
    &\pm \left\{ (K_{z} - K^\prime_{z})^2[(K_{y}+K_{x})^2 + \frac{1}{4}(K_{z}+K^\prime_{z})^2] \right.\\
    &\left. + 4(K_{y}K_{z}- K_{x}K^\prime_{z})(K_{y}K^\prime_{z}- K_{x}K_{z})\sin^2 \frac{k}{2}\right\}^\frac{1}{2}.
\end{aligned}
\end{equation}
The gap is closed at $\vert K_{x} - K_{y}\vert = \sqrt{K_{z}K^\prime_{z}}$. The energy dispersion of the $0$-flux sector can be obtained by the transformation $K^\prime_{z}\to -K^\prime_{z}$. The transition lines where the gap closes are $\vert K_x + K_y\vert = \sqrt{K_{z}K^\prime_{z}}$ and $K_{x}=K_{y}(\vert K_x + K_y\vert > \sqrt{K_{z}K^\prime_{z}})$. The phase diagram of different $\{D_j\}$ sectors of the trimmed ladder model is drawn in Fig.~\ref{fig:pd_realistic}. The solid lines demonstrate the transition lines in the ground state sector, $\{D_j\} = \{+1,-1,,+1,-1,\cdots\}$. The dashed lines represent transition lines in the $0$-flux sector, where $\{D_j\} = \{+1,+1,\cdots\}$. The blue lines reproduce the same phase diagram as $K_{z}=K^\prime_{z}$, while the red lines correspond to the case with $K_z^\prime/K_z = 1/4$. 

One can understand the phase diagram by moving from the $K^\prime_{z}/K_{z}=1$ case to $K^\prime_{z}<K_{z}$ case, the phase transition lines are shifted as the black arrows indicate. When $K^\prime_{z}\to 0$, the two lines in both $\pi$-flux and $0$-flux merge to one line. At that limit, the energy dispersion goes to $\epsilon_{\pm,k} = \sqrt{(K_x+K_y)^2 -4 K_x K_y \sin^2 \frac{k}{2}+ K_z^2/4} \pm K_z/2$. The transition line is $K_x = K_y$. In this limit, there exists macroscopic number of zero modes. Because the $z$-bonds for even number of rungs are missing, the $Z_2$ flux at the missing rungs, $i \Tilde{\gamma}_{2j,\A} \Tilde{\gamma}_{2j,\B}$, have zero energy in the Hamiltonian, resulting in $2^{L/2}$-fold degeneracy. 
The region which is within the two solid lines are of interest, as the model is in the NT phase for the $\pi$-flux sector and is in the SPT phase for the $0$-flux sector. The square root dependence guarantees that even for small $K^\prime_{z}/K_{z}$, for instance, $K^\prime_{z}/K_{z}=0.01$, the region still has a noticeable width of $0.1$ in the phase diagram. 

Note that the engineering of such a material is primarily theoretical in nature. This is because, in order to prevent other symmetry-allowed interactions among $J_{\rm eff}=1/2$ pseudospins, such as Heisenberg and $\Gamma$ interactions \cite{Rousochatzakis_2024,Generic_spin_model_2014}, one must inhibit direct exchange processes, a task that presents significant challenges. Additionally, it is highly probable that the ideal octahedral structure is modified due to the presence of an insulating barrier along the ladder. This would introduce additional interactions which further spoil the exactly solvable problem. 
Additionally, while it has been established that the extended Kitaev ladder model exhibits the same Kitaev phase as the Kitaev point, albeit with small Heisenberg and $\Gamma$ interactions \citep{Jackeli2009,Generic_spin_model_2014,sorensen2024K_gamma_ladder,sorensen2021heart_KJ_ladder}, the identification of Majorana edge modes beyond the exact solvable point remains a question. We defer these inquiries to future studies.

\section{Summary and Discussion\label{sec:summary}}
Identifying and manipulating fractionalized excitations that encode non-Abelian statistics in many body systems are essential for realizing TQC. Examples of systems exhibiting fractionalized excitations and non-Abelian statistics include $p+ip$ superconductor and the Kitaev spin model under the magnetic field. Meanwhile, studies are also focused on 1D $p$-wave superconductors, which host edge Majorana modes in the topological phase. The isolated Majorana modes in wire networks are proposed to encode non-Abelian statistics.\cite{alicea2011non,aasen2016milestones} These studies demonstrate the potential realization of TQC using 1D topological systems. 
Another system that hosts the edge Majoaran modes is the spin$-1/2$ Kitaev ladder model.\cite{Feng2007,inhomo2012,degottardi2011topological,He2013MF,math_paper_2013,math_paper_2015}
However, the boundary Majorana modes in 1D Kitaev ladder model may be absent since an arbitrary superposition of degenerate ground states may annihilate them. 

We explore a way to revive the Majaorana modes in the 1D Kitaev ladder model. We first numerically verify the 4-fold ground state degeneracy for the SPT phase in the open boundary condition as found in previous studies\cite{He_Ladder}, resulting from degenerate flux sectors. We then show that the Majorana modes are indeed absent if we take a superposition of the ground states.
In order to revive the boundary Majorana modes, the PM to select a flux sector is proposed. The PM operator involves a series of spin operators acting on a half of the system, which is hard to implement for today's experiments. We leave the realization of the PM as future work. 
We further show that the flux sector can be manipulated by spin flips on bonds, once the PM chooses a particular flux sector as an initial state, and demonstrate the two Majorana modes become a vacuum or a complex fermion when they are fused together.
The Majorana zero modes in the Kitaev ladder is also related to weak zero modes~\cite{weak_Majorana,Weak_bosonic_mode}. The weak zero modes are the zero modes that only exist within a subspace of the system. In contrast to strong zero modes, the degeneracy exists for all the eigenstates. 
As discussed in Sec.~\ref{sec:manipulating}, the Majorana zero mode in the $0$-flux sector does not exist in the $\pi$-flux sector within a region of the parameter space. Hence, the Majorana zero mode in the $0$-flux sector is a weak zero mode. 
The weak and strong zero modes in the Kitaev ladder would be an interesting problem to study in the future. 
Additionally, we investigate the engineering principles behind creating a Kitaev spin ladder. Given that the Kitaev honeycomb structure is constructed by coupling zig-zag Kitaev chains, the ladder could be actualized by replacing $J_{\rm eff}=1/2$ ions such as RuCl$_3$ with those possessing a filled shell configuration, such as IrCl$_3$ with $d^6$, except for two chains. 
The structure of this ladder is rather different from the original ladder structure because one of the $z$ bonds should be weaker. This results in an alternating $z$-bond strength, a trimmed  ladder. We obtain the phase diagram for the trimmed  ladder and confirm the persistence of Majorana modes. 
Consequently, our method of reviving Majorana fermions remains applicable to an alternative $z$-bond due to the geometry of the trimmed ladder.  

The current study is primarily theoretical. We address the challenges inherent in engineering the Kitaev ladder system, including symmetry-allowed interactions and potential deformations of the octahedral structure. Despite these challenges, our work offers valuable insights into restoring Majorana modes in 1D quantum spin$-1/2$ systems.

\section*{Acknowledgement}
This work is supported by the Natural Sciences and Engineering Research Council of Canada (NSERC)  Discovery Grant No. 2022-04601. H.Y.K acknowledges the support by the Canadian Institute for Advanced Research (CIFAR) and the Canada Research Chairs Program. Computations were performed on the Niagara supercomputer at the SciNet HPC Consortium. SciNet is funded by: the Canada Foundation for Innovation under the auspices of Compute Canada; the Government of Ontario; Ontario Research Fund - Research Excellence; and the University of Toronto.

\bibliography{cites}
\appendix
\section{Jordan-Wigner transformation\label{append:JW}}

The Jordan-Wigner fermion is defined by $f_{j,\mu}^\dagger = \sigma^+_{j,\mu}S(j,\mu)$, where the string operator, $S(j,\mu)$, is defined by 
\begin{equation}
    S(j,\mu) = \left\{
    \begin{aligned}
        &\prod_{\substack{k<j\\ (k,\nu)\in \mathcal{L}_B}} \sigma_{k,\nu}^z,\ &(j,\mu) \in \mathcal{L}_B, \\
        &\prod_{\substack{k<j\\ (k,\nu) \in \mathcal{L}_T}}\sigma_{k,\nu}^z \prod_{(l,\alpha)\in \mathcal{L}_B} \sigma_{l,\alpha},\ &(j,\mu) \in \mathcal{L}_T,
    \end{aligned}
    \right. 
\end{equation}
where $\mathcal{L}_B$ and $\mathcal{L}_T$ represents the set of spins at the bottom ladder and the top ladder, respectively. 
Majorana fermions are defined as
\begin{equation}
\label{eq:JW_trans}
\begin{aligned}
    \gamma_{j,\A} &= f_{j,\A}+ f_{j,\A}^\dagger,&\  \Tilde{\gamma}_{j,\A}&= i(f_{j,\A}^\dagger - f_{j,\A}), \\
    \gamma_{j,\B} &= i(f_{j,\B}^\dagger - f_{j,\B}),&\  \Tilde{\gamma}_{j,\B}&= f_{j,\B}+ f_{j,\B}^\dagger, \\
\end{aligned}
\end{equation}
where they satisfy $\{\gamma_{j,\mu},\gamma_{k,\nu}\}=2\delta_{jk}\delta_{\mu\nu}\mathbbm{1} $, $\{\Tilde{\gamma}_{j,\mu},\Tilde{\gamma}_{k,\nu}\}=2\delta_{jk}\delta_{\mu\nu}\mathbbm{1} $, and $\{\gamma_{j,\mu},\Tilde{\gamma}_{k,\nu}\}=0$. According to the definition in Eq. \ref{eq:JW_trans}, each complex $f$-fermion is represented by two Majorana fermions, thereby preserving the size of the Hilbert space. 

\section{Majorana Boundary mode\label{appendix:boundary_mode}}
To identify the Majorana zero modes in the SPT phase, we review the treatment developed by \citet{degottardi2011topological}. For given parameters in the SPT phase, the Majorana zero modes can be identified as two Majorana operators, $\gamma_\A=\sum_j \alpha_j \gamma_{j,\A}$ and $\gamma_\B = \sum_j \beta_j \gamma_{j,\B}$.
The double degeneracy of the energy spectrum in the topological phase becomes clear by defining a complex fermion operator, $f = \frac{1}{2}(\gamma_A + i \gamma_B)$, such that $f^\dagger f-1/2 = \frac{1}{2}i \gamma_\A \gamma_\B$. Since the $f$ fermion contributes zero energy to the Hamiltonian,
the Hamiltonian resides a $Z_2$ symmetry in the topological phase. The two zero-energy states can be denoted as the eigenstates of the occupation number of $f$ fermion, i.e., $\vert 0 \rangle$ and $\vert 1 \rangle$. By choosing a particular linear combination of the two states, one simultaneously breaks the $Z_2$ symmetry and determines the expectation values of spin or Majorana operators. To elucidate the relation between Majorana mode and the degenerate states, one consider that the operators, $\{\gamma_\A,\gamma_\B,i\gamma_\A \gamma_\B\}$, obey the same commutation relations as the three Pauli matrices. Thus, one can make linear combinations of the state $\vert 0 \rangle$ and $\vert 1 \rangle$ to construct the eigenbasis, $\vert \psi_{\pm,\A}\rangle$ and $\vert \psi_{\pm,\B}\rangle$, of $\gamma_\A$ and $\gamma_\B$ operator, i.e., $\gamma_\A\vert \psi_{\pm,\A}\rangle = \pm \vert \psi_{\pm,\A}\rangle$ and $\gamma_\B\vert \psi_{\pm,\B}\rangle = \pm \vert \psi_{\pm,\B}\rangle$. With these basis, it is straightforward to show that $\langle \psi_{\pm,\A}\vert \gamma_{j,\A}\vert \psi_{\pm,\A}\rangle = \pm\alpha_j$ and $\langle \psi_{\pm,\A}\vert \gamma_{j,\B} \vert \psi_{\pm,\A}\rangle=0$, with the same for $\A \leftrightarrow \B$. As a result, the probability of detecting a Majorana fermion at a specific site for the state $\vert \psi_{\pm,\A}\rangle$ is described by the coefficients, $\vert \alpha_j\vert^2$ for site $(j,\A)$. 

In order to identify the Majorana zero modes, the transfer matrix approach is applied to find the operators $\gamma_{\A}$ and $\gamma_{\B}$. To begin with, one consider the Heisenberg picture with time-dependent operators and time-independent states. The Hamiltonian, Eq.~\ref{eq:K_ladder_fermion}, can be considered as a free fermion Hamiltonian with Bogoliubov quasiparticles. For the fermion in the Hamiltonian with energy $\omega$, the time-dependence is described by $f(t) = f(0) e^{-i\omega t}$. For the $f$-fermion that has the form of $f = \frac{1}{2}(\gamma_\A+ i \gamma_\B)$, the operator $\gamma_\A$ and $\gamma_\B$ also satisfy $\gamma_\A(t) = \gamma_\A(0) e^{-i\omega t}$ and $\gamma_\B(t) = \gamma_\B(0) e^{-i\omega t}$. Substitute the form of $\gamma_\A(t)$ and $\gamma_\B(t)$ into the Heisenberg equation of motion, one obtain the following equations,
\begin{equation}
\label{eq:op_relations}
    \begin{aligned}
        -K_x  \gamma_{j-1,\B}  - K_y \gamma_{j+1,\B} + K_z  D_j  \gamma_{j,\B} &=-i\omega \gamma_{j,\A},\\
        K_x \gamma_{j+1,\A} + K_y \gamma_{j-1,\A} -K_z  D_j \gamma_{j,\A} &=-i\omega \gamma_{j,\B}.
    \end{aligned}
\end{equation}
Because the Majorana zero modes have zero energy, we look at the $\omega =0$ case of Eq.~\ref{eq:op_relations}, which can be solved by real coefficients, $\alpha_j$ and $\beta_j$. Combining Eq.~\ref{eq:op_relations} with the fact that $\{\gamma_\A,\gamma_{j,\A}\}=2\alpha_j$ and $\{\gamma_\B,\gamma_{j,\B}\}=2\beta_j$, one can derive the relations of $\alpha_j$ and $\beta_j$,
\begin{equation}
\label{eq:coeff_relations}
\begin{aligned}
    -K_x \beta_{j-1} - K_y \beta_{j+1} + K_z  D_j \beta_j &= 0,\\
    K_x \alpha_{j+1} + K_y \alpha_{j-1} - K_z  D_j \alpha_{j} &= 0.
\end{aligned}
\end{equation}
These coefficients are normalized  
ensuring that the edge Majorana operators satisfy $\gamma_A^2 = \gamma_B^2 = 1$. To determine the coefficients of the edge Majorana operators, $\gamma_A$ and $\gamma_B$, one starts with an arbitrary boundary condition, such as $\alpha_1 = 1$, $\beta_1 =1$. The remaining coefficients can be determined using Eq.~\ref{eq:coeff_relations}. 

To elucidate the exponential decay of the probability $\vert \alpha_j\vert ^2$, we exactly solve Eq.~\ref{eq:coeff_relations} for $\{D_j\} = \{+1,-1,+1,-1,\cdots\}$, corresponding to the $\pi-$flux sector. From Eq.~\ref{eq:coeff_relations}, the relation between $\beta_j$ is the same as the relation of $\alpha_{L-j}$ for mirror-symmetric $\{D_j\}$. Hence, it is sufficient to exactly solve $\alpha_{j}$'s for this case. 
With $\{D_j\} = \{+1,-1,+1,-1,\cdots\}$, Eq.~\ref{eq:coeff_relations} can be written as
\begin{equation}
    \begin{aligned}
        K_x \alpha_2 + K_z \alpha_1 &=0, \\
        K_x \alpha_3 + K_y \alpha_1 - K_z \alpha_2 &=0, \\ 
        K_x \alpha_4 + K_y \alpha_2 + K_z \alpha_3 &=0, \\
        K_x \alpha_5 + K_y \alpha_3 - K_z \alpha_4 & =0, \\
        &\vdots \\
        K_x \alpha_{L} + K_y \alpha_{L-2} +(-1)^L \alpha_{L-1} & =0. \\
    \end{aligned}
\end{equation}
Starting with the boundary condition such that $\alpha_1 =1$, it follows that $\alpha_2 = -\frac{K_z}{K_x}$. For general integer $n$,  $(\alpha_{2n+1},\alpha_{2n+2})$ are determined solely by $\alpha_{2n-1}, \alpha_{2n}$, through 
\begin{equation}
    \begin{aligned}
        \alpha_{2n+1} &= -\frac{K_y}{K_x} \alpha_{2n-1} + \frac{K_z}{K_x}\alpha_{2n}, \\ 
        \alpha_{2n+2} &= \frac{K_z}{K_x} \frac{K_y}{K_x} \alpha_{2n-1} + \lmbk{-\frac{K_y}{K_x}-\lbk{\frac{K_z}{K_x}}^2} \alpha_{2n}.
    \end{aligned}
\end{equation}
We perform linear combinations of these two equations, such that 
\begin{equation}
    \alpha_{2n+1} + M_{\pm} \alpha_{2n+2} = R_{\pm}\lbk{\alpha_{2n-1}+ M_{\pm} \alpha_{2n}},
\end{equation}
where $M_{\pm} = \frac{1}{2K_y} \lbk{-K_z \pm \sqrt{K_z^2 + 4 K_x K_y}}$ and $R_{\pm} = -\frac{K_y}{K_x}+ M_{\pm} \lbk{\frac{K_z}{K_x}}\lbk{\frac{K_y}{K_x}}$. Hence, the exact general solution of the coefficients $\alpha_j$ can be obtained from
\begin{equation}
\label{eq:coeff_solution}
    \alpha_{2n+1} + M_{\pm} \alpha_{2n+2} = R_{\pm}^n(\alpha_1 + M_{\pm}
    \alpha_2).
\end{equation}
From Eq.~\ref{eq:coeff_solution}, it is clear that the linear combinations of the coefficients, $\alpha_{2n+1}$ and $\alpha_{2n+2}$, exponentially decay or grow as a function of the site index $n$ if and only if $\vert R_{+}\vert>1$ and $\vert R_{-} \vert>1$. 
Hence, a topological invariant, $\nu$, can be defined as 
\begin{equation}
    \nu = -\mathrm{sgn}\lmbk{(\vert R_{+}\vert -1)(\vert R_{-}\vert -1)},
\end{equation}
where $\nu =1$ represent the non-topological phase, and $\nu = -1$ represent the topological phase with edge mode. For $\{D_j\} = \{+1,+1,\cdots\}$, the topological invariant has the following form,
\begin{equation}
    \nu = -\mathrm{sgn}\lbk{(K_y - K_x )^2 - K_z^2}.
\end{equation}
For $\vert K_y - K_x\vert > K_z$, the probability of locating Majorana fermion for $\vert \psi_{+,\A/\B}\rangle$ is exponentially decaying or growing as a function of the site index $j$, indicating the edge Majorana mode. For $\vert K_y-K_x\vert <K_z$, there is no such exponential decay or growth. Hence, the phase is identified as the non-topological phase, with the quantum phase transition lines confirmed as $\vert K_y - K_x \vert = K_z$. These results match with the exact solution.

\begin{figure}
  \centering
    \includegraphics[width = 1.0\linewidth]{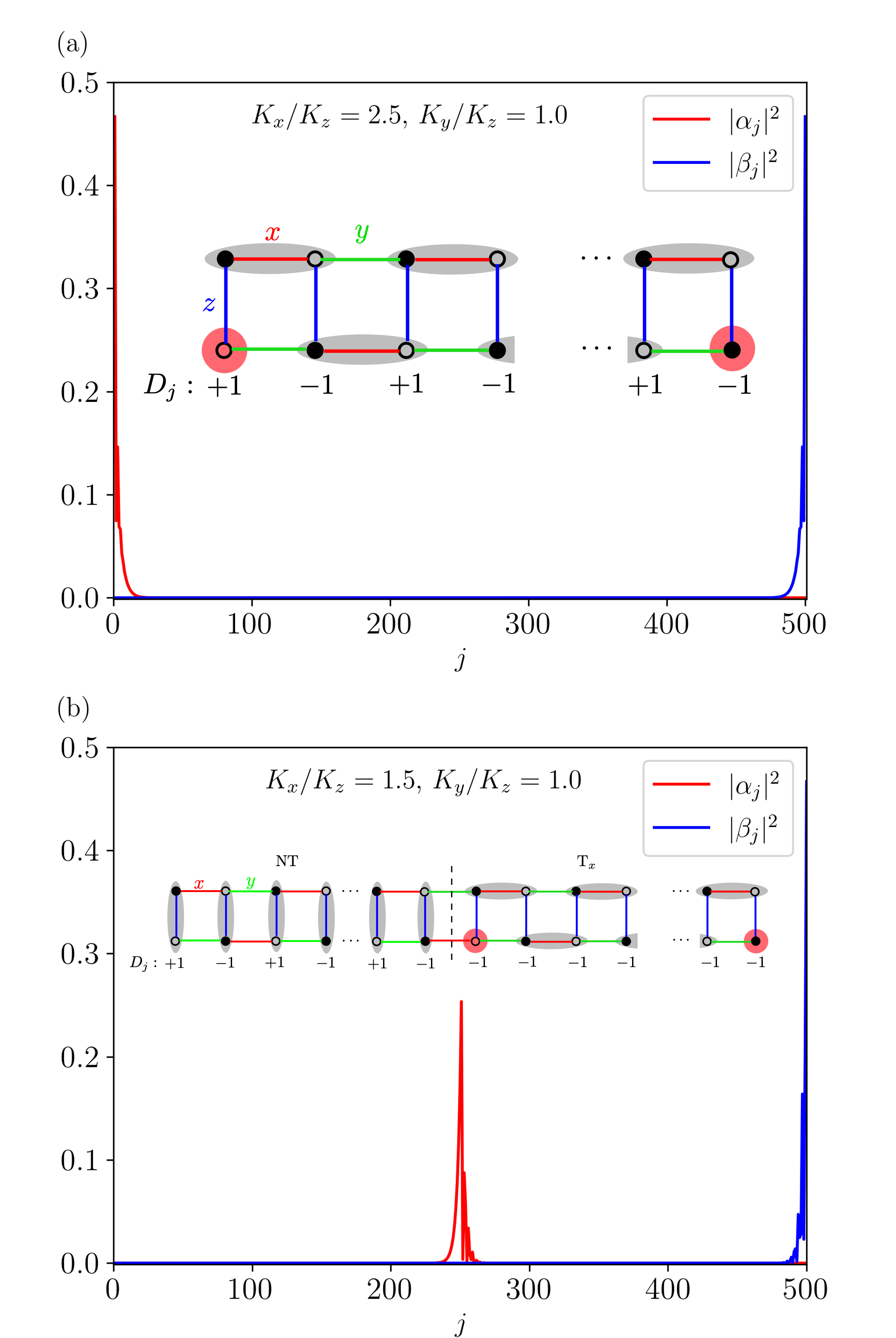}
  \caption{(a) The coefficients of the Majorana zero modes obtained from diagonalizing the fermion Hamiltonian with $K_x = 2.5$, $K_y = 1.0$, $K_z = 1.0$ and $\{D_j\}=\{+1,-1,+1,-1,\cdots\}$. One Majorana zero mode is located at each end of the ladder. The illustration depicts the Majorana fermion correlation in the topological phase, with the red circle indicates isolated Majorana mode. 
  (b) The Majorana zero mode emerges at the interface of the topological phase, $\mathrm{SPT}_x$, and NT phase for the Kitaev ladder system with $K_x = 1.5$, $K_y = 1.0$, $K_z = 1.0$. For $j$ from $1$ to $L/2$, $D_j$ is chosen as $(-1)^j$, which is identified as non-topological phase. For $j$ from $L/2+1$ to $L$, $D_j=-1$, which is identified as $\mathrm{SPT}_x$ phase. The illustration depicts the Majorana zero mode located at the interface of two phases, with red circle represents localized Majorana fermion. 
  }
    \label{fig:isolated_Majorana}
\end{figure}

In order to confirm the Majorana zero mode, we perform numerical simulations with fixed flux sectors and open boundary conditions (OBC). The Hamiltonian is represented as $\hamil = \sum_{j\mu, k \nu}A_{(j\mu,k\nu)}(i \gamma_{j,\mu} \gamma_{k,\nu})$. The quasi-particle excitations are characterized by the linear superposition of Majorana fermions, which is determined by the eigenvectors of the Hermitian matrix $A_{(j\mu,k\nu)}$. 
In Fig.~\ref{fig:isolated_Majorana}(a), we plot $|\alpha_j|^2$ and $|\beta_j|^2$, 
which represent the edge Majorana zero modes on each edge of the system using $N = 1000$, i.e., $L=500$ rungs with $K_x/K_z = 2.5$, $K_y/K_z =1.0$, and $\{D_j\}=\{+1,-1,\cdots\}$. The red circule in the inset indicates the localized Majorana mode, while the grey oval denotes a stronger x-bond in 
the $K_x\gg K_y,K_z$ limit, which is in 
 the $\mathrm{T}_x$ phase.
Note that these coefficients decay exponentially when moving to the bulk of the system, which 
agrees with the analytically obtained $\alpha_j$ and $\beta_j$. 
In addition to the edge of the system, localized Majorana fermion can also emerge at the interface of topological and non-topological phases. To demonstrate this, we diagonalize $A_{(j\mu,k\nu)}$ for a system size of $L=500$, with $K_x/K_z=1.5$, $K_y/K_z=1.0$. The flux sector is chosen as $\{D_j\}=\{+1,-1,\cdots\}$ for $j$ from $1$ to $L/2$, and $\{D_j\}=\{-1,-1,\cdots\}$ for $j$ from $L/2+1$ to $L$. The dashed line in the inset represents the interface of the topological and non-topological phase. The probability of Majorana fermion locating at each site are plotted in Fig. ~\ref{fig:isolated_Majorana}(b). 
The one Majorana mode, which is represented by the red dot near the dashed line, appears at the interface of the non-topological and the topological phases.

\section{Flux degeneracy in NT phase \label{sec:perturbation_theory}}
\begin{figure}
    \centering
    \includegraphics[width=1.0\linewidth]{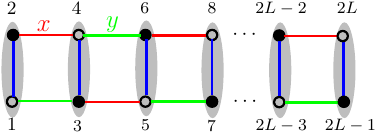}
    \caption{Applying perturbation theory in the limit with $K_z \gg K_x,K_y$. The gray shade indicates that $K_z$ interaction is dominant.}
    \label{fig:perturb_z}
\end{figure}

In order to understand the double degeneracy of the ground states in the non-topological phase, we perform perturbation theory studies. We consider the limit when $K_z \gg K_x,K_y$. The Hamiltonian is 
\begin{equation}
    \hamil = H_0 + V, 
\end{equation}
where $H_0 = \frac{K_z}{4} \sum_{\langle j,k\rangle\in z\text{-bond}} \sigma^z_{j}\sigma^z_{k}$ and $V = \frac{K_x}{4} \sum_{\langle j,k\rangle \in x \text{-bond}} \sigma^x_{j}\sigma^x_{k} + \frac{K_y}{4} \sum_{\langle j,k\rangle\in y \text{-bond}} \sigma^y_{j}\sigma^y_{k}$. We relabel the site as depicted in Fig.~\ref{fig:perturb_z}.
In the zeroth order, there is only $z$-bond Ising interactions. In each vertical bond of the ladder, the quantum state is either $\vert \u\d\rangle_{2j-1,2j}$ or $\vert \d\u\rangle_{2j-1,2j}$, while the states with the same spin alignment, $\vert \u\u\rangle_{2j-1,2j}$ and $\vert \d\d\rangle_{2j-1,2j}$, have the energy $\sim K_z$, which is a much higher energy that we ignore for the construction of the effective Hamiltonian. We denote $\vert \Uparrow\rangle_j = \vert \u\d\rangle_{2j-1,2j}$ and $\vert \Downarrow \rangle_j = \vert \d\u\rangle_{2j-1,2j}$. With these two basis, one can construct the $SU(2)$ algebra with three Pauli matrices, defined as
\begin{equation}
    \begin{aligned}
        \tau_j^x &= \vert \Uparrow_j \rangle \langle \Downarrow_j \vert + \vert \Downarrow_j \rangle \langle \Uparrow_j\vert , \\
        \tau_j^y &= -i\vert \Uparrow_j \rangle \langle \Downarrow_j \vert + i\vert \Downarrow_j \rangle \langle \Uparrow_j\vert, \\
        \tau_j^z &= \vert \Uparrow_j \rangle \langle \Uparrow_j \vert - \vert \Downarrow_j \rangle \langle \Downarrow_j\vert. 
    \end{aligned}
\end{equation}

According to perturbation theory, which is reviewed in ~\cite{kitaev2006anyons}, the effective Hamiltonian has the form of
\begin{equation}
\label{eq:perturbation_theory}
    H_{\mathrm{eff}} =  \Pi_0^\dagger (V+ V G_0^\prime (E_0) V + V G_0^\prime (E_0) V G_0^\prime(E_0) V + \cdots)\Pi_0,
\end{equation}
where $\Pi_0$ is the projector onto the ground state subspace, $G_0^\prime(E_0) = \lbk{(E_0-H_0)^{-1}}^\prime$ is the Green's function that only acts on the excited states and vanishes when acting on ground states. For the first order, we have $H_{\mathrm{eff}}^{(1)} =0$. The first non-vanishing order is the second order, the second order effective Hamiltonian is 
\begin{equation}
    H_{\mathrm{eff}}^{(2)} = - \frac{K_x K_y}{4 K_z} \sum_j \tau_j^y \tau_{j+1}^y,
\end{equation}
which is the Ising model along $\tau^y$ axis. Hence, there exists two fold degenerate ground states, $\vert \psi_1 \rangle$ and $\vert \psi_2 \rangle$, defined as
\begin{equation}
\begin{aligned}
    \vert \psi_1 \rangle &= \vert +_y,+_y,\cdots +_y \rangle= \vert +_y \rangle_1 \vert +_y \rangle _2 \cdots \vert +_y \rangle _L, \\ 
    \vert \psi_2 \rangle &= \vert -_y, -_y, \cdots -_y \rangle = \vert -_y \rangle_1 \vert -_y \rangle _2 \cdots \vert -_y \rangle _L,
\end{aligned}
\end{equation}
where $\vert \pm _y\rangle_j = \vert \Uparrow \rangle_j \pm i \vert \Downarrow \rangle_j$ are two eigenstates of the $\tau_j^y$ operator.
These two states are related through time reversal symmetry, i.e., $T \vert \psi_1 \rangle = \vert\psi_2\rangle$ and vice versa. To see the relation between these two states and $\{D_j\}$ sectors, we express the $D_j$ operators using $\tau$ operators when effectively acting on the ground state subspace, 
\begin{equation}
    D_j = (-1)^{j+1} \tau_j^x \prod_{k\neq j } \tau_k^z. 
\end{equation}
It is straightforward to check that $D_j \vert \psi_1 \rangle = (-1)^{j+1} i \vert \psi_2 \rangle$ and $D_j \vert \psi_2\rangle = (-1)^{j+1} (-i) \vert \psi_1\rangle$. Hence, the two-fold degeneracy can be characterized by the two $\{D_j\}$ sectors, 
\begin{equation}
\begin{aligned}
\vert \psi_1\rangle +i \vert \psi_2 \rangle& \ \mathrm{with} \ \{D_j\} = \{+1,-1,+1,-1,\cdots\}, \\ 
\vert \psi_1\rangle -i \vert \psi_2 \rangle& \ \mathrm{with} \ \{D_j\} = \{-1,+1,-1,+1,\cdots\} .\\ 
\end{aligned}
\end{equation}

\section{Methods of applying spin operators through magnetic field\label{sec:spin_operator}}
In order to change $D_j $ values, 
local spin operators, for instance, $\sigma^x_{j}\sigma^x_{k}$ on $y$-bond, is needed to be applied on the spin system. The local spin operator could be applied through applying a magnetic field on a bond for a short time of $\delta t$. 
To understand the process, let us consider the Hamiltonian $H = \hamil_K+ \hamil_B$, where 
\begin{equation}
    \hamil_B = \frac{\mu_B B\hbar}{2}(\sigma^x_j+ \sigma^x_k )[\theta(t-t_0)-\theta(t-(t_0+\delta t))],
\end{equation}
where $\hamil_K$ is the Hamiltonian for the Kitaev spin ladder, $j,k$ belongs to $y$-bond that connect nearby $D_j$, $\theta(t)$ is the step function with $\theta(t)=0$ for $t<0$ and $\theta(t)=1$ for $t>0$. When considering the time evolution of the state with Schr\"{o}dinger equation, the state evolves from $t_0$ to $t_0+\delta t$ as
\begin{equation}
    \vert \psi(t_0+\delta t)\rangle = e^{-i(\hamil_K +\hamil_B)\delta t/\hbar} \vert \psi(t_0)\rangle   .
\end{equation}
In order to prevent the system to enter another phase,
$\delta t $ need to be controlled to satisfy $\frac{E\delta t}{\hbar}\ll 1$, where $E$ is the energy of the state. With these approximation, the time evolution can be approximated by 
\begin{equation}
    \vert \psi(t_0+\delta t)\rangle \simeq  e^{-i\hamil_B\delta t/\hbar} \vert \psi(t_0)\rangle .
\end{equation}
Hence, to effectively apply the operator $\sigma_j^x \sigma_k^x$, the magnetic field should satisfy 
\begin{equation}
\frac{\mu_B B\delta t}{2} =  \frac{\pi}{2},
\end{equation}
such that $e^{(-i \sigma^x \pi/2)} = -i \sigma^x$. 

Hence, by applying a strong magnetic field with a short period of time on a local bond, one can effectively apply the local spin operator to change the $D_j$ sector. Although the topological nature of the Hamiltonian can be changed through applying these local spin operators, the dynamics of the quantum state needs to be further studied.

\end{document}